\documentclass[aps,twocolumn,showpacs,amsmath,amssymb,floatfix]{revtex4}
\usepackage{epsf}

\begin{document}

\title {Condensation energy in strongly coupled superconductors.}  

\author{Robert Haslinger$^1$, Andrey V. Chubukov$^2$}

\affiliation{$^1$ Los Alamos National Laboratory, Los Alamos, NM 87545}
\affiliation{$^2$ Department of Physics, University of Wisconsin, 
Madison, WI 53706}   
\date{\today}   
 
\begin{abstract}   
We consider the condensation energy, $E_c$,
 in superconductors where the pairing is electronic in origin and is 
mediated by a collective bosonic mode.
 We use magnetically-mediated superconductivity as an example, and show
  that for spin-fermion couplings $\lambda \sim 1$,
 the physics is qualitatively different from the BCS theory 
 as the condensation energy 
 results from the feedback on 
 spin excitations, while the electronic contribution to $E_c$ is positive
 due to an ``undressing'' feedback on the fermions.
 The same feedback effect accounts for the gain of the  
 the kinetic energy  at strong couplings.
\end{abstract}  
\maketitle

 Undertanding the origin of the condensation energy is an 
important step towards identifying the mechanism of 
high temperature superconductivity in the cuprates. 
In a BCS superconductor, the 
condensation energy - the energy gained upon entering the 
superconducting state, is $E_c^{BCS} = - N_f \Delta^2/2$, 
where $\Delta$ is the superconducting gap and 
$N_f$ is the fermionic density of states. \cite{bcs} This decrease in the total
energy upon pairing results from a fine 
competition between an increased kinetic energy and a decreased  
potential energy, both of which are much larger than $E_c$. 
 For optimally doped  $YBCO$ with $N_f \sim 10^{-3} meV^{-1}$, 
  and $\Delta \sim 30 meV$, the BCS formula yields   
$E_c \sim 5K$ in 
 reasonable agreement with  $E_c \approx 0.12 k_BT_c \sim 10 K$ that 
  Loram et al~\cite{loram} extracted from  specific heat measurements.
 However, at smaller dopings, $\Delta$ increases while the measured $E_c$
 decreases, in clear disagreement with the BCS theory. 

In this paper, we present a computation of the condensation energy for
 a strongly coupled superconductor. We argue that at strong coupling, 
 the relation between $E_c$ and $\Delta$ is 
different from BCS theory, and is
 consistent with experimental trends in the underdoped cuprates. 
 We show that the 
%numerical 
agreement with the BSC theory 
at optimal doping is largely coincidental as at this doping range, 
strong coupling effects are already dominant.
 
Some earlier works have already suggested that
 the gain in the condensation energy in the cuprates may be due to
 a non BCS physics. 
 Scalapino and White~\cite{scal_white} conjectured that at strong coupling, 
the dominant contribution to the condensation energy comes from a feedback 
effect on the magnetic excitations of the system.
On the other hand,  Norman~{\it et al} 
 argued~\cite{mike1} 
that the condensation energy  likely has an electronic 
 origin, and is driven by a gain in the kinetic energy which at strong
  coupling is negative (in contrast to BCS theory) 
 because of a strong ``undressing'' of fermions which 
  bear a greater resemblance to free particles in the superconducting state 
than they do in the normal state.  Similar ideas were expressed 
 by Hirsh and Marsiglio~\cite{hirsh}. A different idea, related 
 to the lowering of the Coulomb energy in the superconducting state
  has been proposed by 
 Leggett~\cite{leggett}. 

In this communication we argue that these apparently  disparate
viewpoints are in fact consistent with each other, and describe the 
same strong coupling physics.
 In the
 superconducting state, spin decay into fermions is forbidden
  at energies smaller than $2\Delta$.
This  simultaneously gives rise to two effects. First, 
 the spin propagator develops 
 an excitonic (resonance) peak at $\omega_{res} < 2\Delta$. 
  The energy released by the creation of an exciton results in a 
  gain in the magnetic part of the condensation energy. 
 Secondly, the fermions cannot  decay 
 until their frequency exceeds $\Delta + \omega_{res}$ (this is  the
 magnetic analog of the Holstein effect).
 The elimination of fermionic scattering at low
  frequencies implies that  the fermionic self-energy $\Sigma (\omega)$ 
  in the superconducting state is reduced compared to 
that in the normal state. 
 This effect lowers the kinetic energy (see below) and at  
 strong coupling overcomes the effect of particle-hole mixing that increases
$E_{kin}$ in a BCS superconductor. 

The above reasoning is quite general and can be applied to various 
superconducting mechanisms. A more subtle issue is to
 explain the reduction of $E_c$ in the  underdoped regime, and
 why $E_c \sim 10K$ at optimal doping despite the fact that by all accounts, 
strong coupling effects are already large. 
Below we present  an explicit computation  of $E_{c}$ assuming that the 
 pairing is due to spin-fluctuation exchange 
%and is described by  the spin-fermion model.
% which he obtained  by using the effective 
%functional method  of  Luttinger and Ward~\cite{lutt_ward}. 
 We show that our theoretical $E_c$ agrees
 with the data both in magnitude and doping dependence.
  We view this agreement as  
support for the spin-fluctuation scenario for the cuprates. 

 The condensation energy $E_c$ is the difference between the
 grand free energies in the superconducting and normal states~\cite{mu}
%\begin{equation}
$E_c =  \Omega_{s} - \Omega_{n}$.
%\end{equation}
 We obtained $E_c$   by modifying  
 the Eliashberg formula for phonon 
superconductors~\cite{eliash,lutt_ward} to the spin case.
This approach implies that  $E_c$ is dominated
 by momenta  near the hot spots  where  the 
 momentum dependence of the $d-$wave gap can be neglected
 except for 
 the sign change between ``hot'' regions. We will  discuss 
 the validity of the Eliashberg theory for the spin case 
 after we present the results.  
In the Eliashberg theory, $E_c$   can be represented as the
sum of two parts $E_c = \delta \Omega_{el} + \delta \Omega_{spin}$, where
\begin{eqnarray*}
\Omega_{el} &=& -T \sum_m \int \frac{d^2 k}{(2\pi)^2}   
(\log[\epsilon _{\mathbf{k}}^{2}+\widetilde{\Sigma} 
_{\omega_m}^{2}+\Phi^2_{\omega_m}] \nonumber\\
 &-& i \Sigma_{\omega_m} G_{\omega_m} (k) +i
\Phi_{\omega_m} F_{\omega_m} (k) ) 
\end{eqnarray*}
\begin{eqnarray}
\Omega_{spin} &=& \frac{3}{2} T \sum_m \int \frac{d^2 q}{(2\pi)^2} 
[\log (\chi^{-1} (q, \omega_m))  
 \nonumber\\ &+& \Pi_{\omega_m} \chi (q, \omega_m)/\chi(Q,0)].
\label{eliash0}
\end{eqnarray} 
Here $G_(\omega_m)$ and $F(\omega_m)$ are the real and anomalous
Greens functions given by  $G_{\omega _{m}} (k) =-(\epsilon _{\mathbf{k}}+i
\widetilde{\Sigma}_{\omega_m})/(\epsilon _{\mathbf{k}}^{2}+\widetilde{\Sigma}
_{\omega_m}^{2}+\Phi^2_{\omega_m})$ and $F_{\omega _{m}} (k) =
i\Phi (\omega _{m})/
(\epsilon _{\mathbf{k}}^{2}+\widetilde{\Sigma}_{\omega_m}^{2}+
\Phi^{2}_{\omega_m}) $, where $\epsilon_{\mathbf{k}} = v_F (k - k_F)$, and
 $\widetilde{\Sigma}_{\omega_m} = \omega_m + \Sigma_{\omega_m}$. 
The conventionally defined pairing gap 
$\Delta_{\omega_m}$ is the ratio of 
the anomalous vertex and the self-energy: $\Delta_{\omega_m} = \Phi_{\omega_m} \omega_m/{\tilde \Sigma}_{\omega_m}$. In the second term, 
 $\chi \left( \mathbf{q}, \omega \right)$ is 
the dynamical spin susceptibility 
 related to the spin polarization operator  
$\Pi_{\omega_m}$ by  
 $\chi \left( \mathbf{q}, \omega_m \right) = \chi (Q,0)/(1+\xi ^{2}\left( 
\mathbf{q-Q}\right) ^{2}-\Pi_{\omega_m})$ where 
$\chi (Q,0)$ is the static staggered susceptibility.
 The factor of 3 in $\Omega_{spin}$ is due to the spin summation.
%it is  absent in the phonon case.

The term $\delta\Omega_{el} = \Omega_{el}^{sc} - \Omega_{el}^{n}$ 
accounts explicitly for the appearance of the anomalous pairing vertex 
$\Phi_{\omega_n}$, {\it and} for the feedback changes to the 
fermionic self-energy. 
 The term $\delta\Omega_{spin} = \Omega_{spin}^{sc} - \Omega_{spin}^{n}$
accounts for changes to the spin propagator via the
 changes to the spin polarization operator $\Pi_{\omega_m}$.
 This term is almost negligible for 
 phonon superconductors~\cite{bardeen_stephen}.  It is however
 important for magnetic superconductors where the bosonic mode that mediates the superconductivity is by itself affected
 by the pairing.

The fermionic self-energy $\Sigma_{\omega_m}$, the pairing  
vertex $\Phi_{\omega_m}$, and the spin polarization 
 operator $\Pi_{\omega_m}$ 
are the solutions of three coupled Eliashberg equations
~\cite{acf}
\begin{eqnarray}
\Sigma_{\omega _{m}} &=&  \pi \lambda T\sum_{n}
~\frac{\widetilde{\Sigma}_{\omega_n}}{\sqrt{\widetilde{%
\Sigma}^2_{\omega_n}+\Phi^2_{\omega_n}}} 
~\frac{1}{(1 -\Pi_{\omega_m - \omega_n})^{1/2}},\nonumber\\ 
\Phi_{\omega _{m}} &=& \lambda \pi T\sum_{n}
\frac{\Phi_{\omega_n}}{
\sqrt{\tilde{\Sigma }^2_{\omega_n}+\Phi^2_{\omega_n}}}~
\frac{1}{(1 -\Pi_{\omega_m - \omega_n})^{1/2}}, 
\nonumber \\ 
\Pi _{\omega_m} &=&
  \frac{4 \lambda^2}{\bar {\omega}}  \pi T \sum_{n}[-1
 \nonumber\\ &+& 
~\frac{\widetilde{\Sigma}_{\omega_n} \widetilde{\Sigma}_{\omega_n + \omega_m} 
 + \Phi_{\omega_n} \Phi_{\omega_n + \omega_m} 
}{\sqrt{\widetilde{
\Sigma}^2_{\omega_n}+\Phi^2_{\omega_n}} \sqrt{\widetilde{
\Sigma}^2_{\omega_n + \omega_m}+\Phi^2_{\omega_n + \omega_m}}}].
\label{eliash1}
\end{eqnarray} 

Eqs. (\ref{eliash1}) contain only two inputs: 
the overall energy scale $\bar \omega$ that is set 
by the spin-fermion interaction,
and the dimensionless spin-fermion 
coupling $\lambda \propto \xi$ that diverges as 
 the system approaches the antiferromagnetic instability.
Physically, the energy scale $\bar \omega$   is the 
 ultimate upper  cutoff for the strong coupling behavior 
 ($\Sigma_{\omega_m} < \omega_m$ for $\omega_m > {\bar \omega}_m$), while 
 dimensionless $\lambda$ can be represented as 
 the ratio $(2\lambda)^2 = {\bar \omega}/\omega_{sf}$ 
of ${\bar \omega}$
 and another typical scale $\omega_{sf}$ that sets the 
 upper boundary of the Fermi-liquid behavior in the  normal state.

The evolution of the condensation energy with 
 $\lambda$ can be understood as follows. 
At small $\lambda$,  i.e., at weak coupling, 
 ${\bar \omega} < \omega_{sf}$, 
and the system behaves as a
 conventional Fermi liquid.
 In this limit, 
 the pairing potential is static, i.e., the 
condensation energy is entirely electronic and BCS like.
At strong couplings, i.e. $\lambda \geq 1$,  ${\bar \omega} > \omega_{sf}$,
there is 
a frequency range between $\omega_{sf}$ and ${\bar \omega}$ 
within which the system behaves as
 a non Fermi liquid.  Previous studies demonstrated~\cite{acf}
 that in this regime both the onset of the 
 pairing instability and the magnitude of $\Delta_{\omega}$ at $T=0$ are 
 determined by ${\bar \omega}$,  and hence 
the condensation energy results predominantly from 
 fermions located in the non-Fermi liquid frequency range. For these fermions,
 retardation effects not included in the BCS theory become dominant,
 hence one expects strong deviations from the BCS form of $E_c$.
Comparison to experiment~\cite{mike1,acf} yields $\lambda \sim 1.5-2$ 
at optimal doping. This doping already corresponds to a strong coupling 
(${\bar \omega} \sim (10-16) \omega_{sf}$).

As $\Sigma$ and $\Phi$ depend only on $\omega$,
the momentum integration in Eqs. (\ref{eliash0}) can be 
performed explicitly and yields
\begin{eqnarray} 
\delta \Omega_{el} &=& - N_f \pi T \sum_m (\sqrt{\widetilde{
\Sigma}^2_{s,\omega_m}+\Phi^2_{\omega_m}} - |{\widetilde \Sigma}_{n,\omega_m}|
 \nonumber\\
 &+& |\omega_m| \frac{|{\widetilde \Sigma}_{s,\omega_m}| - \sqrt{\widetilde{
\Sigma}^2_{s,\omega_m}+\Phi^2_{\omega_m}}}{\sqrt{\widetilde{
\Sigma}^2_{s,\omega_m}+\Phi^2_{\omega_m}}})\\
\label{kin2el}
\delta\Omega_{spin} &=& - N_s  \pi T \sum_m \nonumber \\
&&\left({\tilde \Pi}_{s, \omega_m} - {\tilde \Pi}_{n,\omega_m} + 
 \omega_{sf} \log{\frac{\omega_{sf}  - {\tilde \Pi}_{s,\omega_m}}
{\omega_{sf} - {\tilde \Pi}_{n, \omega_m}}}\right) 
\label{kin2spin}
\end{eqnarray}
where  ${\tilde \Pi}_{\omega_m} = \omega_{sf} \Pi_{\omega_m}$.
 In the normal state, ${\tilde \Pi}_{\omega_m} = - |\omega_m|$.  
We introduce $N_f$ via $\int d^2k/(4\pi^2) = N_f \int d\epsilon_k$ and 
an analogous spin ``density of states''
 $N_s = (8/3\pi^2) {\bar \omega}/v^2_F$, where $v_F$ is the Fermi velocity.
%We assume that both $N_f$ and $v_F$ are  direction independent.
One can  make sure that for $\lambda \geq 1$, 
the frequency sums in both terms yield 
$O(({\bar \omega})^2)$, hence 
$\delta \Omega_{el} \sim N_f {\bar \omega}^2$, while 
$\delta \Omega_{spin} \sim N_s {\bar \omega}^2$.

\begin{figure}[tbp]
\epsfxsize= 3.2in
\begin{center}
\leavevmode
\epsffile{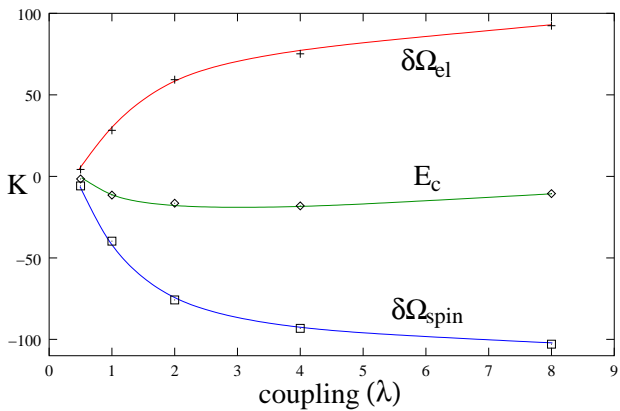}
\end{center}
\vspace{-0.75cm}
\caption{Electronic ($\delta\Omega_{el}$),  
spin ($\delta\Omega_{spin}$), and total ($E_c$) condensation energy per
unit cell at $T=0$ for various couplings
$\lambda$. The lines
are a guide for the eye.  The sum of $\delta\Omega_{el}$ and $\delta\Omega_{spin}$
produces a total condensation energy which is {\it negative}.
 We used $N_f = 1 st/eV$, and $N_s \sim 0.17 st/eV$ as explained
in the text.}
\vspace{-0.5cm}
\label{fig:1}
\end{figure}

In Fig.\ref{fig:1} we present the results for the 
electronic and spin contributions to the condensation energy 
for different values of the coupling $\lambda$.
 We used previously obtained solutions of 
the Eliashberg equations at $T \approx 0$ 
in real frequencies, and
 related $\delta \Omega$ with 
 $\Sigma_{\omega}$, $\Phi_{\omega}$ and $\Pi_{\omega}$ along the real 
frequency axis using the  spectral representation. 

There are three striking features of Fig.\ref{fig:1}. First, 
the electronic contribution to the condensation 
energy is {\it positive}.  As in the BCS limit  the
electronic condensation energy is negative and equal to $-N_f \Delta^2/2$,
this implies that the electronic condensation energy 
changes sign at a rather small $\lambda \sim 0.4$, 
and is positive for all $\lambda \geq 1/2$ presented 
in the figure. 
Second,  for all $\lambda$ shown,
the spin part $\delta \Omega_{spin}$ 
is negative. Third, at large $\lambda$, 
%both the spin and the electronic 
 the two parts of the condensation energy nearly saturate.

We next have to find  a relation between $N_s$ and $N_f$. 
 We use a trick employed by Bardeen and 
Stephen~\cite{bardeen_stephen} and introduce a ``mixed''
 self-energy $\Sigma_{ns}$ that has the 
 normal state form, {\it but} with the spin propagator from  
the superconducting state. 
We then introduce the functional 
\begin{equation}
Y_{\omega_m} (k) =  i  \left(G_{n,\omega_m} (k)
 \Sigma_{n,\omega_m} -G_{ns, \omega_m} (k) \Sigma_{ns,\omega_m}\right) 
\nonumber
\end{equation}
and compute $Y = T \sum_m \int \frac{d^2 k}{(2\pi)^2}~Y_{k, \omega_m}$
 in two ways -- first, explicitly integrating over $\epsilon_k$, and second, 
reexpressing the fermionic self-energy 
via the spin polarization operator and then integrating over the 
bosonic momentum. 
The explicit integration yields 
\begin{equation}
Y = \pi T N_f \sum_m  \left( |\Sigma_{n,\omega_m}| - 
|\Sigma_{ns,\omega_m}|\right)  
\label{y1}
\end{equation}
Expressing $\Sigma$ in terms of $\Pi$, 
and using the fact that  the polarization operator made out of $G_{ns}$ is 
 precisely $\Pi_n (\omega)$ as the latter  depends on the 
 functional form of the Green's function but not 
 on the magnitude of the self-energy~\cite{acf}, 
we find after simple algebra
\begin{equation}
Y =  - \pi N_s \sum_m {\tilde \Pi}_{n, \omega_m} 
\log{\frac{\omega_{sf}  -{\tilde \Pi}_{s,\omega_m}}
{\omega_{sf} - {\tilde \Pi}_{n, \omega_m}}}
\label{y2}
\end{equation}  
Comparing (\ref{y1}) and (\ref{y2}) we see that $N_f$ and $N_s$ are indeed
 related. We evaluated the ratio for various $\lambda$ and found that 
with small variations, $N_f/N_s \approx 5.9$ for all $\lambda \geq 0.5$. 

In Fig.~\ref{fig:1} we present the result for the 
full condensation energy $E_c$ using this ratio  of $N_f/N_s$.
To set the overall scale, we 
 adopt a commonly used estimate $N_f = 1st/eV$~\cite{abanov}.
 We emphasize that
changing $N_f$ will only change the overall
scale and not the functional form of $E_c(\lambda)$.
We see from  the figure that the 
 total condensation energy is  negative, as it indeed should be in
  a superconductor, but its value is rather small due to a 
  substantial cancellation 
 between the spin and electronic contributions to $E_c$. 

\begin{figure}[tpb]
\epsfxsize= 3.2in
\begin{center}
\leavevmode
\epsffile{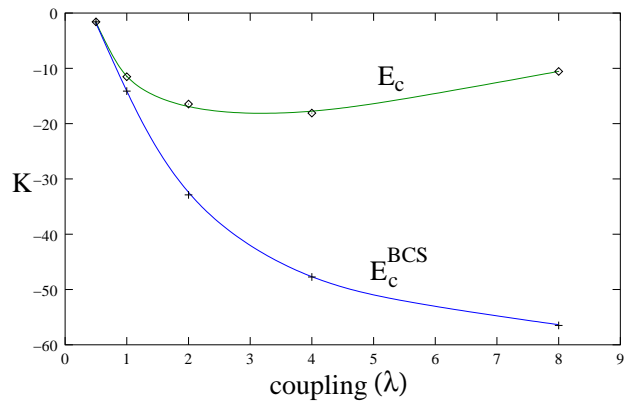}
\end{center}
\vspace{-0.75cm}
\caption{Total condensation energy $E_c$ compared
with the BCS result $E_c^{BCS}$ at $T=0$ for various
couplings $\lambda$. We used $N_f = 1 st/eV$ and $N_f/N_s \sim 5.9$ as 
 explained in the text. Observe that the  BCS condensation energy 
 increases monotonically as the coupling gets larger, while 
 the actual 
condensation energy  flattens at 
 $\lambda \sim 2$ and  
decreases slightly at large couplings.}
\vspace{-0.5cm}
\label{fig:2}
\end{figure}

We also see that  the condensation energy flattens 
at $\lambda \sim 2$, and decreases
 at large couplings despite the fact that the pairing gap 
 increases monotonically with $\lambda$~\cite{acf}.
The decrease in $E_c$ can  be understood as a reflection of the fact that 
as $\lambda$ increases the pairing process 
more and more involves the exchange of  classical, on-shell bosons.  Due to
energy conservation, such bosons  
 can not lead to a gain in $E_c$.~\cite{artem}.  This behavior is 
very counterintuitive from a BCS perspective, 
 where the condensation energy scales with $\Delta^2$.  
 To further emphasize this point, we  plot in Fig.\ref{fig:2} our $E_c$ and  
 the BCS condensation energy $-N_f\Delta^2/2$ with  the same 
$\Delta$ and $N_f$.
We clearly see that for $\lambda \geq 1$, corresponding
 to optimally doped and  underdoped cuprates, BCS theory yields
  qualitatively incorrect results for $E_c$. 

\begin{figure}[tbp]
\epsfxsize= 3.2in
\begin{center}
\leavevmode
\epsffile{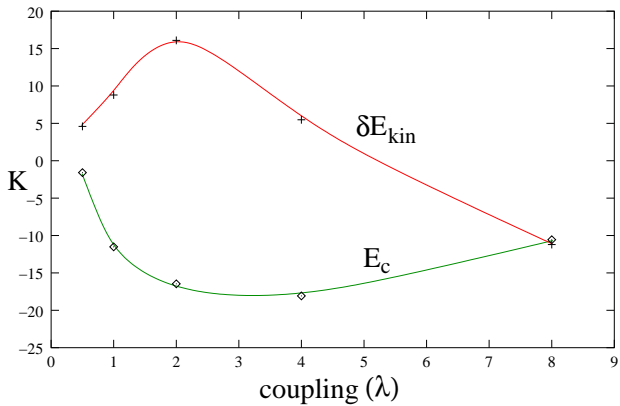}
\end{center}
\vspace{-0.75cm}
\caption{Kinetic energy $\delta E_{kin}$ compared with total condensation energy
$E_c$ at $T=0$ for various couplings $\lambda$.
The parameters are the same as in Fig.  2.  The kinetic energy change 
is positive at low couplings, but negative at high coupling.}
\vspace{-0.5cm}
\label{fig:3}
\end{figure}

We next present the results for the 
 change in the kinetic energy when the system enters 
the superconducting state.
The conventionally defined kinetic-energy for an interacting fermionic system 
 is  
\begin{equation}
E_{kin} = 2  T\sum_m  \int \frac{d^2 k }{(2\pi)^2}~ \epsilon_k~ G_{\omega_m} (k) 
\label{kin}
\end{equation}
where $G_{\omega_m} (k)$ is the full fermionic Green's function 
that contains the self-energy. Integrating over momentum and subtracting the
normal state result from $E_{kin}$ in a superconductor 
we obtain
\begin{equation}
\delta E_{kin} = 2N_f \pi T \sum_m \sqrt{\widetilde{
\Sigma}^2_{s,\omega_m}+\Phi^2_{\omega_m}} - |{\widetilde \Sigma}_{n,\omega_m}|
\label{simpkin}
\end{equation}
  In the BCS limit, $\Phi_{\omega_n} = \Delta$, 
$\Sigma =0$, ${\widetilde \Sigma}_{\omega_n} = \omega_n$ and 
$\delta E_{kin}=  2N_f \pi T \sum_m \sqrt{
\omega_m^2+\Delta^2} - |{\omega_m}|$ which is obviously positive.
The results at finite $\lambda$ are presented in Fig~\ref{fig:3}.
 We used the same computational procedure as before. 
At low couplings the kinetic energy is positive, as one naively
expects.  At larger $\lambda$, however, the kinetic energy passes through a
maximum at $\lambda \sim 2$ and then becomes negative at 
large $\lambda$. This implies, as we anticipated,
 that at strong coupling, the 
 lowering of $E_{kin}$ via the change in the self-energy 
due to  the ``undressing'' of fermions   
overcomes the effect of particle-hole mixing that increases $E_{kin}$.
   This behaviour is very similar to that
obtained by Norman et. al. \cite{mike1}.

 As the first term in $\delta\Omega_{el}$ is equal to $-\delta E_{kin}/2$
 (eqs. \ref{kin2spin} and \ref{simpkin}),
 one can indeed argue that 
 the condensation energy at large couplings  is at least partly 
 driven by the lowering of the kinetic energy.  However, 
 a comparison of Figs. \ref{fig:1} and \ref{fig:3} shows that 
 this is just another way to interpret strong coupling effects that 
 affect {\it both} the fermionic and bosonic propagators via mutual feedback.
 
We emphasize that the
 calculated value of $E_c$ is small, only $\sim 15 K$ around optimal doping. 
This is rather remarkable as all typical energies in the problem are much 
higher, i.e ${\bar \omega} \sim (2.5-3) \hspace{0.1cm} 10^3 K$. 
This small value of $E_c$ is partly 
due to small prefactors, but is also the result of substantial
 cancellation between the spin and electronic contributions to $E_c$. 
Loram et al~\cite{loram} extracted  $E_c \approx 0.12 k_BT_c \sim 10 K$ 
from the jump of the  specific heat at $T_c$. Our result is
 is very close to this value.
 
 Loram {\it et al} found that the condensation energy drops as one moves into 
the underdoped regime~\cite{loram}. We also found
 a decrease in  $E_c$ at large couplings, 
although not as spectacular as in Loram's experiments.
This is due to the strong 
 dispersion of the bosonic propagator assumed in the spin-fermion model.
 A weaker spin excitation dispersion, 
 as suggested by neutron experiments ~\cite{keimer}, 
 should result in a stronger reduction of $E_c$. 
 
  Finally, we discuss the accuracy of the Eliashberg approach and also 
clarify what we mean by strong coupling.
As discussed previously ~\cite{acf}, 
the validity of Eliashberg theory is related to the 
development of  an ``effective'' Migdal 
 theorem at strong coupling allowing vertex corrections to be
 neglected.  As spin fluctuations become overdamped
 due to spin-fermion exchange  
they become slow compared to fermions (just as phonons are slow compared 
 to electrons). This effective Migdal theorem emerges if the momentum 
 integration is  confined to the vicinity of the hot spots which  in 
 turn requires   
 ${\bar \omega}$ to be smaller than the fermionic bandwidth $W$. 
This defines a ``universal'' 
strong coupling limit: $\lambda \geq 1, {\bar \omega} < W$.
 The neglect of the $k-$dependence of the $d-$wave gap and 
 of the self-energy along the Fermi surface 
(i.e., the approximation of $N_f$ by a constant) is an 
 $O(1)$ approximation even when ${\bar \omega} \ll W$, but  
 these two $k-$dependences are totally irrelevant from the 
physics perspective and can be safely omitted~\cite{acf}.
Furthermore, even when ${\bar \omega} > W$,
 the pairing problem 
 does not change qualitatively 
except that typical momenta measured from a hot spot 
 are now  $O(p_F)$, and  $N_s$ and $N_f$ are no longer universally related.
 In this regime, the spin part of the condensation energy
 prevails, and  scales 
 as $W^2/{\bar \omega} \sim J$, as
predicted by Scalapino and White ~\cite{scal_white}. 

To conclude, we have shown that at strong spin-fermion coupling, 
the  behavior of the condensation energy
differs strongly from the BCS prediction.  Instead of getting larger
as $\Delta$  increases, as predicted by BCS, the total
condensation energy flattens off and then decreases. 
This decrease appears to be
correlated to a change in sign of the kinetic energy, another non-BCS
type behavior. We also found that
 the spin and charge contributions to $E_c$ are of comparable 
 strength and opposite
sign, and negative $E_c$ 
 results from  a delicate balance between the two.
This behavior has no analog for phonon superconductors, where 
the feedback on bosons is a minor effect. 

We thank Ar. Abanov, 
D. Pines, J. Loram, M. Norman and D. van-der-Marel
 for useful discussions. This research was supported by
  NSF DMR-9979749 (A. Ch)
and  by DR Project 200153, 
 and the Department
 of Energy, under contract W-7405-ENG-36.
 (R.H.) 
%\vspace{-0.5cm}


\begin{thebibliography}{99}
\bibitem{mu}
This implies that the chemical potential 
 does not change between the normal and superconducting states. 
 For continuous low-energy theories, this is the case 
 if the theory is free from ultraviolet divergencies.
Our computational procedure
 that involves a transformation to the real axis 
 is ultraviolet convergent.
\end{thebibliography}
\end{document}